## Article information

### Article title

*Time-of-flight photon spectroscopy for scanning tunneling microscopy luminescence*

### Authors

Lebin Yu[1], Jiří Doležal[2], Maximilian Rödel[2], Amandeep Sagwal[3,4], Benjamin Frölich[1], Martin Švec[2,4], Fabian Donat Natterer[1*]

### Affiliations

1. *Department of Physics, University of Zurich, Winterthurerstrasse 190, CH-8057 Zurich, Switzerland*
2. *Institute of Organic Chemistry and Biochemistry, Czech Academy of Sciences; Flemingovo náměstí 542/2. Praha 6 CZ16000, Czech Republic*
3. *Faculty of Mathematics and Physics, Charles University; Ke Karlovu 3, CZ12116 Praha 2, Czech Republic*
4. *Institute of Physics, Czech Academy of Sciences, Cukrovarnická 10/112, Praha 6 CZ16200, Czech Republic*

### Corresponding author's email address and Twitter handle

*fabian.natterer@physik.uzh.ch*

### Abstract

*We build and commission a time-of-flight photon spectrometer (TOFS) for scanning tunneling microscopy luminescence (STML). We obtain the spectrum by exploiting the wavelength dependent refractive index of a long dispersive optical fiber that converts photon arrival times into wavelength information; blue photons are delayed more than red photons. The setup uses a pulsed excitation source, either laser flashes or voltage pulses, to launch the photons from the junction into a single photon detector. The TOFS calibration can be performed and transferred to the STML setup from a separate benchtop experiment using pulsed light sources with known wavelength. We verify the TOFS during STML operation by simultaneously recording luminescence from the Ag-Ag(111) plasmon using a conventional grating spectrometer. Our experiments show that the TOFS is a straightforward and cost-effective addition to existing STML setups with good performance in the near-infrared range. The TOFS is compatible with a drop-in replacement of the single photon detector such as a superconducting nanowire single photon detector or even bolometers that may expand the useful spectral range beyond the abilities of current STML setups.*

- *Time of flight spectroscopy with single photon detector*
- *Performance potential in long wavelength range*
- *Cost effective spectrometer*

## Graphical abstract

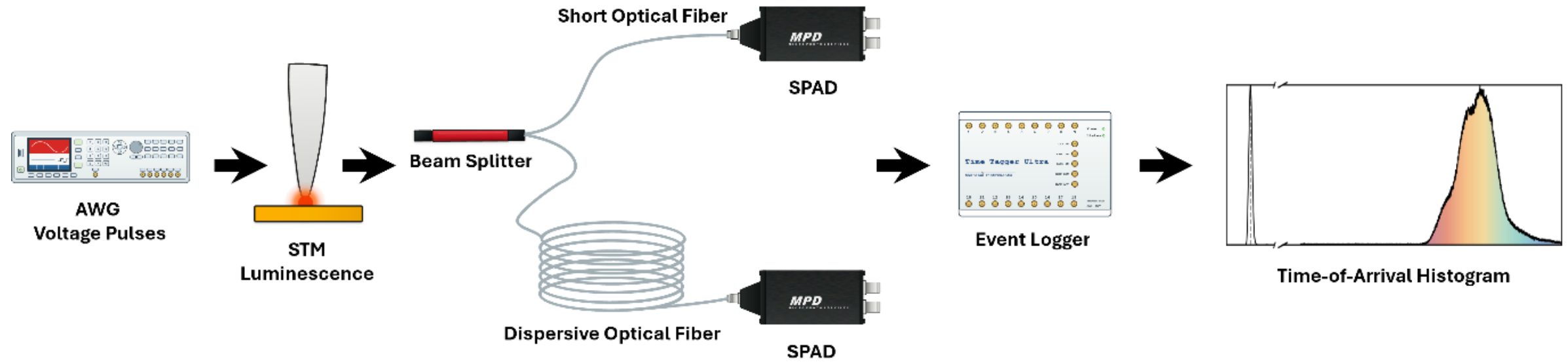

## Specifications table

| | |
|---|---|
| **Subject area** | Physics and Astronomy |
| **More specific subject area** | *Photon Spectroscopy* |
| **Name of your method** | *Time of flight spectrometer for STML* |
| **Name and reference of original method** | *Whitten, W. B., & Ross, H. H. (1979). Fiber optic waveguides for time-of-flight optical spectrometry. Analytical Chemistry, 51(3), 417–419.* *https://doi.org/10.1021/ac50039a022* *Loumaigne, M., Vasanthakumar, P., Richard, A., & Débarre, A. (2012). Time-of-Flight Photon Spectroscopy: A Simple Scheme To Monitor Simultaneously Spectral and Temporal Fluctuations of Emission on Single Nanoparticles. ACS Nano, 6(12), 10512–10523. https://doi.org/10.1021/nn304842c* |
| **Resource availability** | *NA* |

## Background

*Light emission from microscopic sources, such as scanning tunneling microscopy luminescence (STML) of single molecules, defect centers, and plasmonic nanocavities, is typically extremely weak, requiring sensitive single-photon detectors that are often operated in cooled environments. Extracting spectral information under such low-photon flux conditions adds further experimental complexity, because conventional grating spectrometers rely on precise dispersion across a one-dimensional array of single-photon detectors, so that dark counts and noise are added for every wavelength-bin. A time-of-flight spectrometer (TOFS), on the other hand, obtains spectral information with a single photon-detector and a pulsed light source. The wavelength dependent group refractive index of a long optical fiber disperses photons in time, and the resulting arrival-time histogram encodes spectral information.*

*The TOFS concept offers several advantages. First, the dark count rate of the single-photon detector is distributed over the entire spectrum rather than contributing to each wavelength-bin individually, improving the effective detection threshold. Second, the detector can be readily replaced with more advanced sensors, such as superconducting nanowire single photon detectors*[1,2] *or bolometers, provided that they offer sufficient timing resolution. The time domain approach allows straightforward wavelength tracking and filtering just by monitoring photon flux at selected arrival times. Unlike earlier TOFS implementations that relied on pulsed lasers and ensemble measurements*[3–5]*, our approach uses a pulsed voltage source, enabling an accessible integration into existing STML setups. The TOFS is cost-effective and can extend studies into unconventional wavelength ranges, such as in the near infrared (NIR). It can therefore support photon correlation analyses and precise wavelength filtering or monitoring, for instance in optically detected magnetic resonance.*

## Method details

*As the name suggests, a time-of-flight spectrometer (TOFS) extracts spectral information from photon flight-times, dispersing light in time rather than in space as in a grating spectrometer. The TOFS described here exploits the wavelength dependent group-refractive index* $n_g(\lambda)$ *of a long optical fiber to separate photons of different colors by their transit times* $t_{transit}(\lambda) = n_g(\lambda)L/c$*, where* $L$ *is the fiber length and* $c$ *is the speed of light. When the dispersion is monotonic, a unique mapping between* $t_{transit}(\lambda)$ *and wavelength exists on either side of the zero-dispersion wavelength (ZDW)*[3,5]*. Many optically dense materials exhibit such monotonic dispersion*[6]. Accurate wavelength assignment requires precise knowledge of both *fiber dispersion and fiber length. We therefore calibrate these quantities following the approach of Loumaigne*[5]*. For our demonstration, we use fused silica with a ZDW near* $1.3\ \mu m$*, providing unique wavelength assignment from the visible range up to the ZDW and from ZDW into the NIR.*

Figure 1 *shows the experimental setup including calibration, which uses a wavelength tunable pulsed laser (NKT SuperK EXTREME), two single photon avalanche diodes (SPAD, Micro Photon Devices $PD-050-CTE-FC,* $35\ ps$ *timing resolution, 48% peak quantum efficiency at* $550\ nm$*), and a time-tagger with* $42\ ps$ *timing accuracy (Swabian Instruments Time-Tagger Ultra). We couple the laser output into a 50:50 fiber splitter (ThorLabs TM200R5F1A); the short branch directly feeds one SPAD, and the nominally* $1000\ m$ *long branch feeds the second SPAD. For each wavelength, we record* $t_{transit}(\lambda)$ *for both branches twice, swapping the SPADs between the short and the long arms to account for system delays in time-tagger, laser trigger, internal detector delays, and from long electrical cables, allowing us to define an exact photon insertion time into the long fiber.*

*The histogram in* Figure 1 *and fiber calibration in* Figure 2 *show how transit times vary with wavelength. To convert* $t_{transit}(\lambda)$ *into a group-refractive index, we use the known group refractive index* $n_g(850\ nm) = 1.496$ *(Thorlabs Spec sheet) as a reference, yielding a fitted fiber length of* $L = (1000.62 \pm 0.11)\ m$*.* Figure 2 *shows the resulting wavelength dependent group-refractive index from 485 nm to 845 nm.*

*The ratio of photon count rates in short and long fiber branches provides the wavelength dependent attenuation, shown in* Figure 2*. As expected for fused silica, attenuation is significantly higher at shorter wavelengths (*$\sim 13\ dB$ *at* $500\ nm$*) and decreases towards the NIR (*$< 2.5\ dB$ *at* $800\ nm$*), emphasizing the utility of fused silica for long-wavelength spectroscopy.*

*The TOFS method is compatible with different STML variants*[7]*, including ellipsoid/parabolic mirror collection*[8–13]*, in-vacuum lens collection*[8,14–16]*, and direct fiber-coupling*[11,17]*. We describe here the integration of TOFS into STML setups using vacuum lenses*[16,18]*. Similar to the calibration setup, the TOFS requires a pulsed excitation source, a single photon detector, a long optical fiber, and a time-tagger. We use an arbitrary waveform generator (AWG, Keysight 33600A, Keysight 81160A)*[19] *to produce voltage pulses that excite the Ag-Ag(111) plasmon, thus stroboscopically emitting light. The AWG provides the timing reference for the TOFS by sending a synchronized TTL trigger to the time-tagger for each excitation pulse (start-time, $t_{start}$), while the SPAD outputs report the photon arrival times (stop-time, $t_{stop}$). The TOFS is connected on the ambient air side of the STML setup without requiring any permanent modifications and is operated simultaneously with a conventional grating spectrometer for comparison.*

*To illustrate the importance of fiber length, we estimate typical photon fluxes encountered in STML-TOFS experiments. Taking a $1\ nA$ current setpoint used in constant bias STML* but now applying *a voltage pulse of $t_{pulse} = 1\ ns$ duration for the TOFS excitation, tunnels approximately $6.25$ electrons per pulse, each emitting $6.25 \times 10^{-5}$ to $6.25 \times 10^{-3}$ photons*[20] *due to the poor STML quantum efficiency (QE). For a $1000\ m$ fiber and a repetition rate equal to the inverse of the transit time (here $200\ kHz$), the photon flux would accordingly range from $12.5$ to $1250\ photons/s$. Such low fluxes require detectors with very low dark count rate or shorter fibers, although shorter fibers reduce spectral resolution.*

*Alternatively, the TOFS signal-to-noise can be improved by increasing the pulse repetition rate, provided the arrival time spread satisfies $t_{spread} < t_{transit}$. This condition can be met intrinsically by the emission characteristic of the junction or by inserting a suitable optical bandpass filter. When using a filter or working with limited spectral bandwidth, the pulse repetition rate can be increased to the inverse of the total temporal window required to avoid overlap between successive photon-arrival distributions: $1/(t_{spread} + t_{pulse})$, or $1/(\Delta t_{bandpass} + t_{pulse})$. In our example, this allows repetition rates up to $8\ MHz$ (assuming $t_{spread} + t_{pulse} = 125\ ns$), increasing the photon flux by an order of magnitude ($125$ to $1.25 \times 10^{4}\ photons/s$ for the possible range of QE), comfortably above the SPAD dark count rate. We refer to this operating regime as pulse loading of the fiber. Pulse loading was used in all our TOFS validation measurements.*

Figure 1 *shows the raw arrival time histogram of $\Delta t = t_{stop} - t_{start}$ for the calibrated $L = 1000.6\ m$ long fiber, measured by exciting the Ag-Ag(111) plasmon using voltage pulses of $5\ V$ amplitude, $1\ ns$ duration (1.5 ns risetimes), and an $8\ MHz$ repetition rate (pulse loading the fiber). The histogram is centered around $5.01\ \mu s$, corresponding to the mean transit time $t_{transit}(\lambda) = n_g(\lambda)L/c$. The spread $t_{spread} = t_{blue} - t_{red} = 60\ ns$ signals the wavelength dependent dispersion, with red photons arriving earlier and blue photons later. This rather small spread highlights the need for high timing precision to resolve the spectral structure; shorter fibers reduce $t_{spread}$ and thus yield worse spectral resolution (see* Figure S2*).*

## Method validation

*To validate the TOFS, we simultaneously compare its output with that of a conventional Czerny-Turner grating spectrograph (Andor Kymera 328i with CCD Newton 920). A 50:50 cube beam-splitter directs part of the STML emission into the long fiber via a lens ($f = 30\ mm$ achromatic doublet, uncoated) both mounted on a custom 5-degrees of freedom mechanical mount with a cage system (Thorlabs), also shown in the schematics of* Figure 1*. The cage system allows insertion of optical filters, such as a notch filter (Semrock NF03-633E) suppressing $633 \pm 13\ nm$ to introduce a known spectral feature into the broadband Ag-Ag(111) plasmon emission. Additional filters can be employed to isolate spectral regions below and above the zero-dispersion wavelength (ZDW), which for our fiber is $\sim 1350\ nm$.*

Figure 3 *shows the Ag-Ag(111) plasmon that we excite with voltage pulses, as measured with TOFS, overlaid with the simultaneously recorded spectrum from the grating spectrograph, and the uncalibrated notch filter transmission graph (taken from Semrock spec sheets). The notch filter produces a clear suppression around 633 nm in both spectra, with sharp edges and slightly shifted and rounded long wavelength edge compared to the nominal filter characteristic. The rounding at long wavelength stems from the slight diverging beam in the cage system, which we verified in a separate measurement of the notch filter at angled incidence. The TOFS uses only the calibration parameters determined earlier; intensity corrections account for fiber attenuation and SPAD quantum efficiency. Attenuation correction enhances the*

*short wavelengths region, while detector efficiency correction increases the weight above and below* $550\ nm$ *without shifting spectral features. On the one hand, the good agreement of the notch filter spectral feature in grating spectrograph and the optical fiber that we calibrated in a table-top experiment shows that the TOFS calibration can be transferred to other STML systems. On the other hand, the TOFS timing offsets could also be calibrated using a known spectral feature created by an optical filter or a molecular emission line.*

Figure 4 *shows the plasmonic emission spectra, again measured simultaneously with TOFS and grating spectrometer, at pump voltages ranging from* $1$ *to* $5\ V$ *added on top of a 1 V DC voltage. The luminescence onset clearly shifts to shorter wavelengths at higher voltages, consistent with DC measurements of* Martín-Jiménez *et al.*[21]*. A mismatch between the expected onset and the applied voltage suggests an uncalibrated AWG output or voltage transients, which may also contribute to the long wavelength spectral ripples observed in* Figure 3.

## Limitations

*There are limitations of the TOFS that merit discussion:*

***A)*** *One important requirement is a well-defined start time for the TOFS, to exactly know* $t_{transit}$*. Voltage transients and reflections caused by impedance mismatches of cable faults, however, cause time-dependent distortions, leading to ringing and more generally to broadened pulse shapes in the junction. Since these limitations have been previously described, there are effective mitigation strategies available, such as pulse shaping for STML as has been shown previously*[22] *and recently*[13]*.*

***B)*** *The ZDW introduces ambiguity because wavelengths on either side can share the same group-velocity. An optical filter must therefore block either the shorter or longer wavelength side to restore unique mapping. For fused silica (ZDW* $\sim 1.3\ \mu m$*), long and short pass filters can separate the UV-VIS-NIR or VIS-NIR range.*

***C)*** *Fiber attenuation can significantly reduce photon transmission. In fused silica and shown in* Figure 2*, attenuation exceeds* $10\ dB/km$ *below 520 nm (*$\sim 10\%$ *transmission) but remains below* $5\ dB/km$ *up to* $\sim 2.4\ \mu m$*. Users must balance desired wavelength range, fiber length, fiber material, spectral resolution, and coupling efficiency. Recent STML investigations report large collection efficiencies*[13] *that may compensate for large fiber losses. Note that we have not yet optimized the count rate in the TOFS because of the simultaneous comparison with the grating spectrometer. If the signal was directed only into the TOFS branch, we would get an immediate photon doubling and by additionally choosing a 1:9 or 1:99 fiber splitter, we will reach about 4-fold larger count rate compared to the data shown in this work. The SPAD of the short branch will still receive sufficient photons for the timing reference.*

***D)*** *The SPAD used here has a peak quantum efficiency below* $50\%$ *at* $550\ nm$ *and* $< 5\%$ *below* $400\ nm$ *and above* $900\ nm$*. Short-wavelength suppression is particularly problematic due to higher fiber attenuation. Conversely, the excellent long-wavelength transmission of fused silica can offset low detector efficiency in the NIR. This advantage could be further boosted by the use of alternative photo detectors, such as superconducting nanowire single photon detectors (SNSPD)*[23]*. These novel detectors have advantages making them interesting drop-in replacement options for SPAD in the TOFS setup*[1,2]*. For instance, they have an outstanding quantum efficiency near unity in the NIR range, they do not manifest afterpulsing or post-detection electroluminescence, and they have excellent time resolution*[24]*, essential for TOFS.*

***E)*** *In our demonstration, the Ag-Ag(111) plasmonic emission has negligible excited state lifetime but molecular emitters may fluoresce with substantial emission delays, artificially "slowing" down photons in the TOFS picture. A short fiber and optical filter can be used to measure the lifetime and deconvolve it from the TOFS spectrum*[5]*. Alternatively, increasing the fiber length reduces the notable impact of lifetime related delays.*

## CRediT author statement

***Lebin Yu:*** *Conceptualization, Methodology, Validation, Investigation, Writing – review & editing.* **Jiří Doležal:** *Validation.* **Maximilian Rödel:** *Validation.* **Amandeep Sagwal:** *Validation.* ***Benjamin Frölich:*** *Validation.* ***Martin*** **Švec:** *Validation, Conceptualization, Supervision, Funding acquisition.* ***Fabian D Natterer:*** *Conceptualization, Validation, Investigation, Writing – original draft, Supervision, Funding acquisition.*

## Acknowledgments

*Funding: This work was supported by the Swiss National Science Foundation [200021_200639, PP00P2_211014]. J.D., A.S. and M.S. acknowledge the support from the CzechNanoLab Research Infrastructure supported by MEYS CR (LM2023051).*

## Declaration of interests

☒ The authors declare that they have no known competing financial interests or personal relationships that could have appeared to influence the work reported in this paper.

☐ The authors declare the following financial interests/personal relationships which may be considered as potential competing interests:

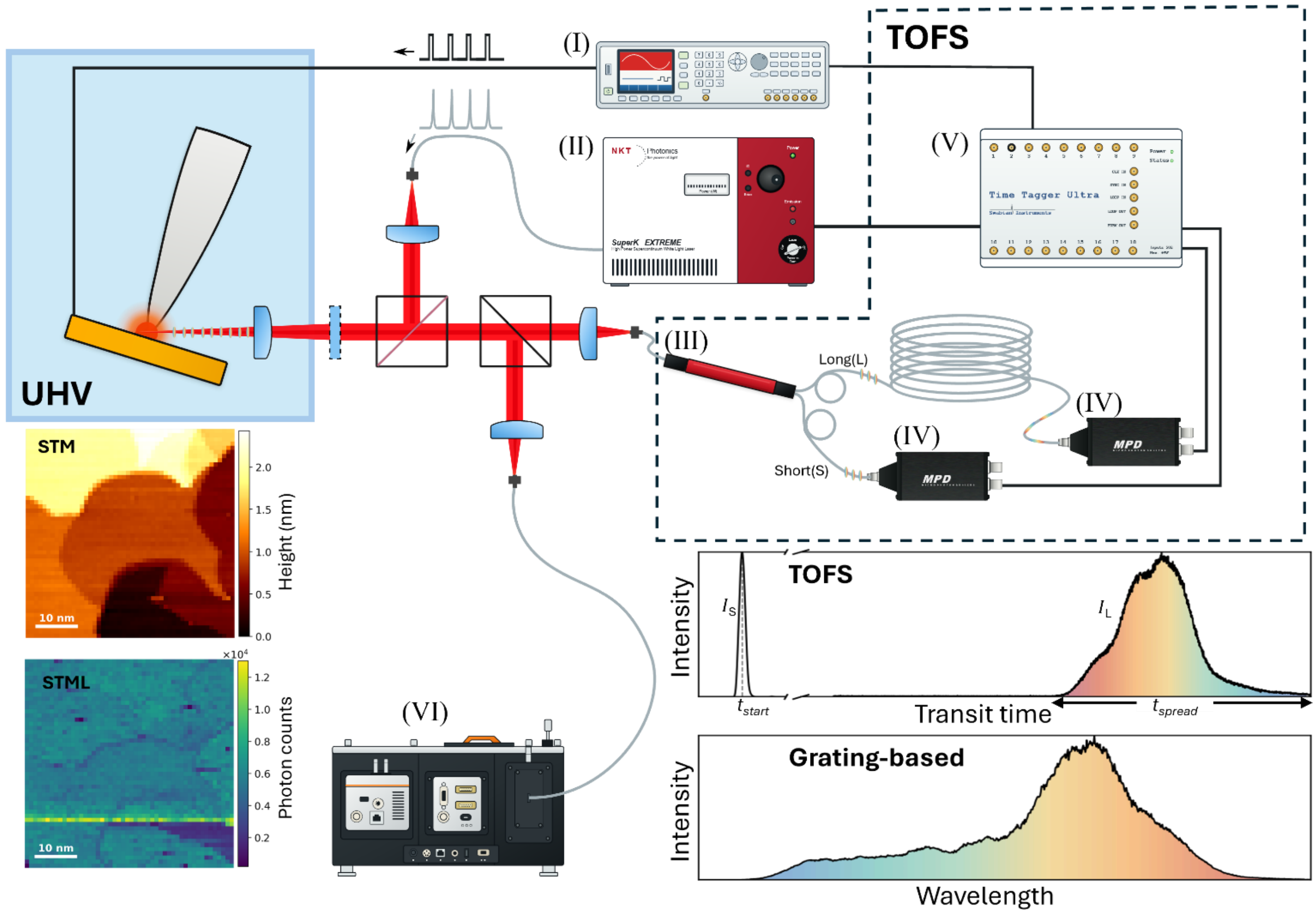

*Figure 1: Sketch of the experimental setup used for time-of-flight spectroscopy (TOFS) of scanning tunneling microscopy luminescence (STML). An arbitrary waveform generator, AWG* (I)*, is used to trigger electroluminescence from the STM junction in an ultrahigh-vacuum (UHV) chamber, while a wavelength tunable laser* (II) *is used for TOFS calibration. The TOFS setup, indicated by the dashed box, consists of a 50:50 fiber splitter* (III)*, which divides the photon stream into two optical paths. One branch is coupled through a short fiber to a single-photon avalanche diode, SPAD,* (IV) *to define the temporal reference. The other branch propagates through a 1 km-long optical fiber before being detected by a second SPAD* (IV)*. The timestamps of the electrical pulses from the AWG, the laser trigger, and the SPAD detection events are collected and recorded by a time tagger* (V)*. In parallel with the TOFS measurement, a conventional grating-based spectrometer* (VI) *is used to benchmark the performance of the TOFS system. The lower-left panel shows the STM topography and corresponding STML signal acquired on Ag(111) at a bias voltage of* $V = 2.5\,V$*, a tunneling current of* $I = 10\,nA$*, and a dwell time of 500 ms/pixel. The lower-right panel shows representative data formats comparing two different spectroscopic methods. The grating-based spectrometer records photon intensity as a function of wavelength, whereas TOFS records photon intensity as a function of photon time of flight through the long optical fiber.*

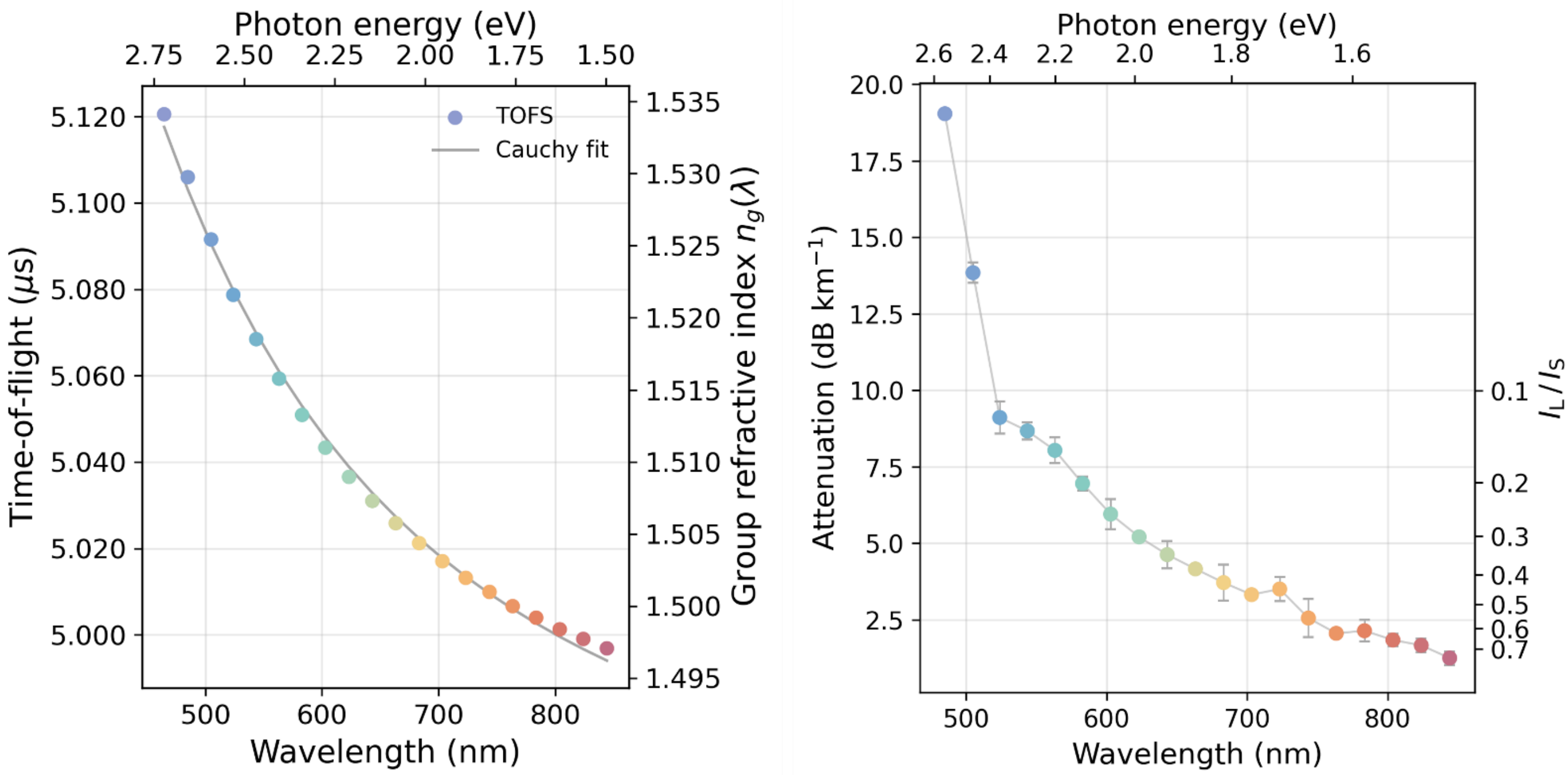

*Figure 2: Dispersion and attenuation characterization of a $1000\ m$ optical fiber (Thorlabs GIF625-1000) used for TOFS. (a) TOFS arrival times measured as a function of wavelength using a tunable pulsed laser, SPADs, and a grating spectrometer for comparison (see* Figure 1*). The solid grey line represents a two-term Cauchy fit describing the wavelength-dependent group refractive index $n_g(\lambda)$, with $n_g(850\ nm) = 1.496$ as constraint. The corresponding $n_g(\lambda)$ values are shown on the right axis. (b) Measured attenuation coefficient of the same fiber as a function of wavelength (circles), showing a pronounced increase in loss at shorter wavelengths. The secondary axis indicates the equivalent transmission ratio $I_L/I_S$.*

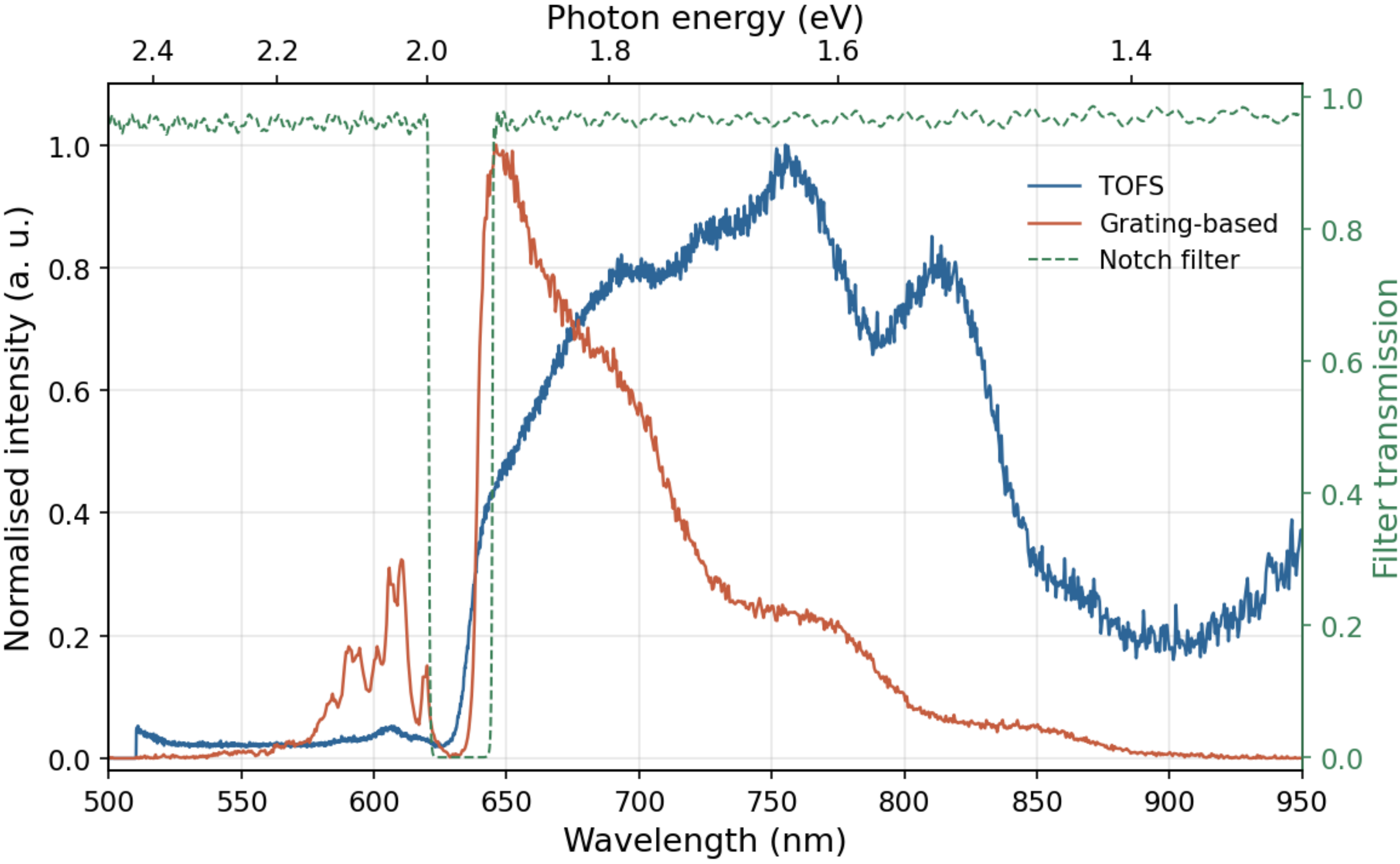

*Figure 3: Simultaneously obtained spectra from TOFS (blue) and grating spectrometer (orange) of Ag-Ag(111) plasmon with an added notch filter ($633 \pm 13\ nm$) creating a known spectral feature. The dashed green curve (right axis) indicates the uncalibrated transmission of the notch filter used in the optical path, with vertical dashed lines marking its nominal cutoffs. The TOFS data reproduces the overall spectral profile and extends into the near-infrared, while differences in intensity arise from wavelength-dependent detection efficiency and fiber attenuation. The agreement around the filter edges highlights accurate timing and wavelength calibration and establishes consistency between TOFS and conventional grating-based spectroscopy. The intensity of the TOFS data was corrected using the wavelength dependent attenuation and detector efficiency. These corrections do not shift spectral information but only change the relative intensity emphasis.*

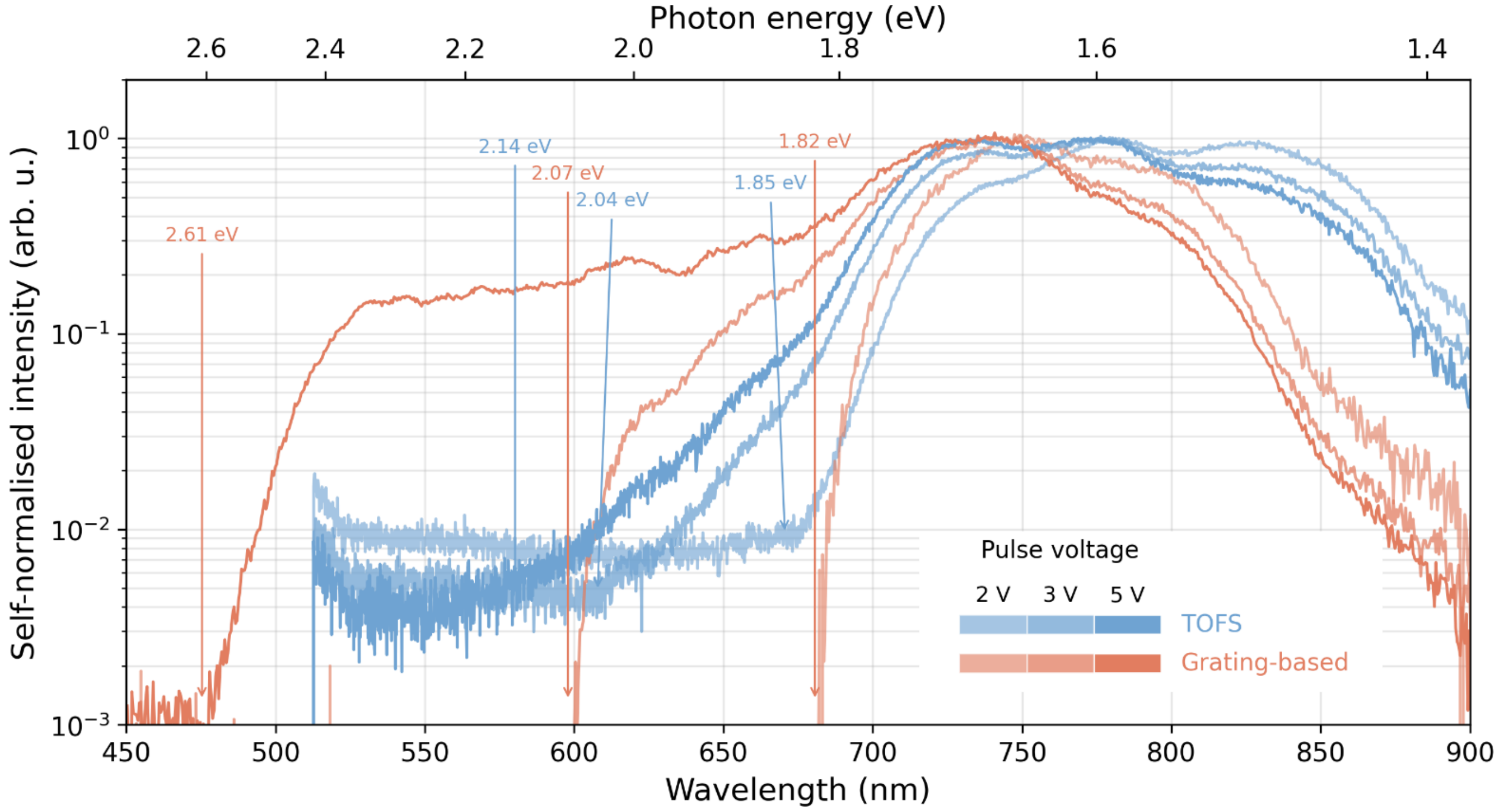

*Figure 4: Voltage dependent photon spectroscopy of Ag-Ag(111) plasmon, as simultaneously seen by the grating spectrograph (orange) and TOFS. The shade encodes the excitation voltage added on top of a 1 V DC bias voltage, and the vertical arrows mark the onset of luminescence at short wavelengths. The discrepancy of the onset threshold compared to the excitation voltage is attributed to the uncorrected AWG calibration and to voltage transients from impedance mismatches.*

**Supplementary material**

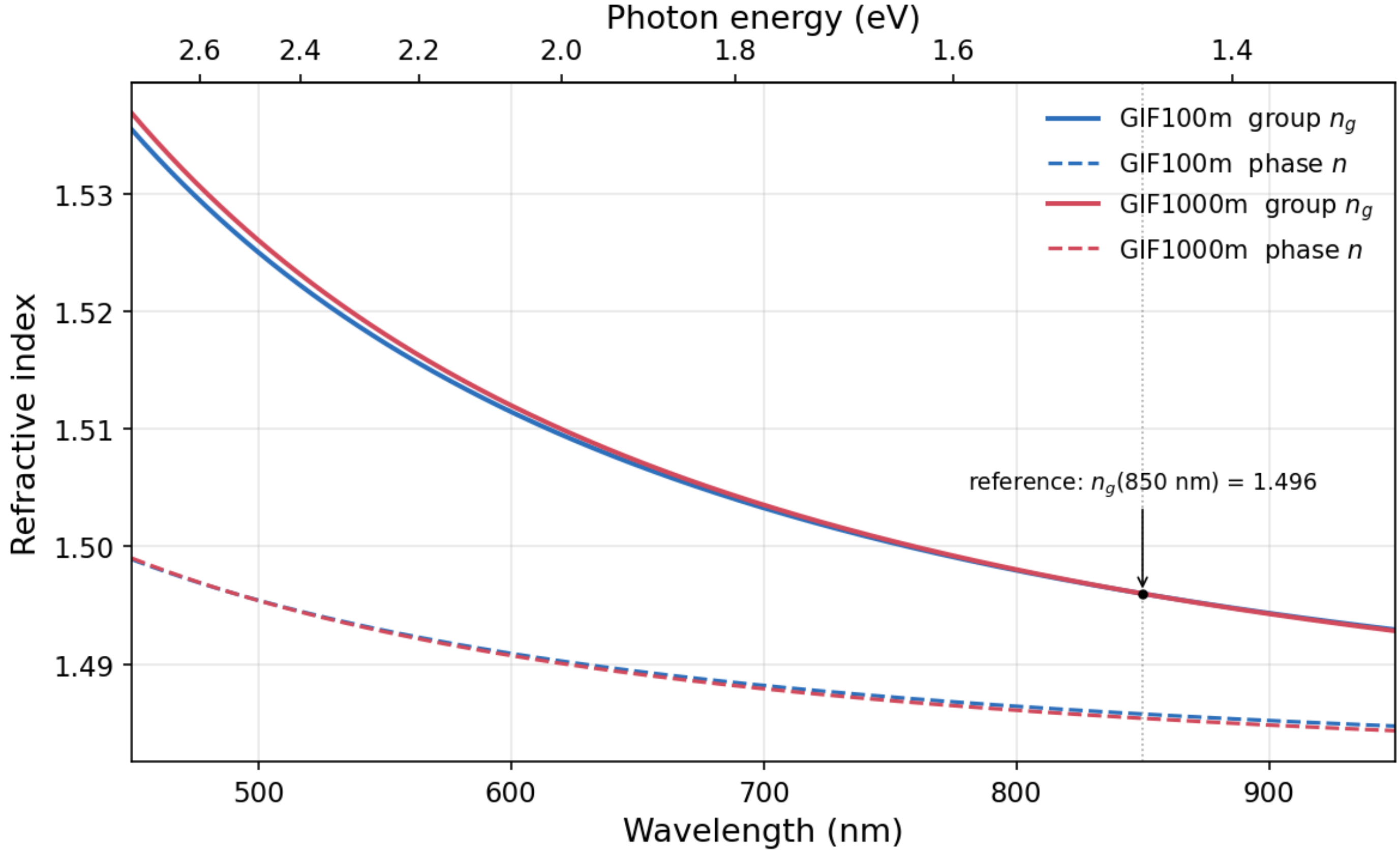

*Figure S1: Refractive index, independently measured for* $100\ m$ *and* $1000\ m$ *long fibers, shows good agreement. The black dot corresponds to the reference refractive index [*$n_g(850\ nm) = 1.496$*], used as a fitting constraint, as explained in* Figure 2*. Dashed lines show phase refractive indices.*

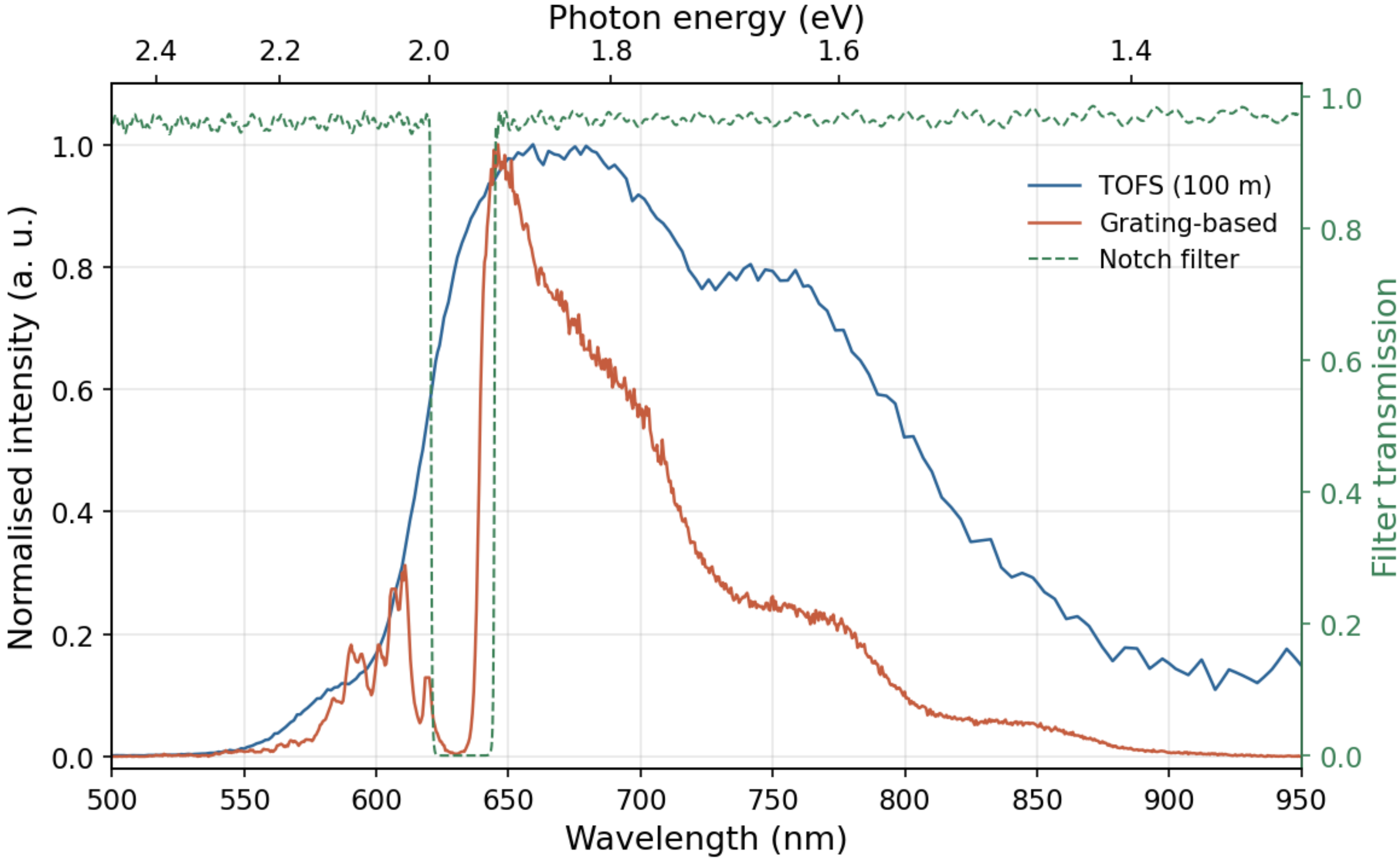

*Figure S2: As in* Figure 3*, simultaneously obtained spectra from TOFS (blue) and grating spectrometer (orange) of Ag(111) plasmon with an added notch filter ($633 \pm 13$ $nm$) for a $100$ $m$ long TOFS. The dashed green curve (right axis) indicates the uncalibrated transmission of the notch filter used in the optical path, with vertical dashed lines marking its nominal cutoffs. Unlike for the $1000$ $m$ case, the resolution of the short fiber is insufficient to detect the spectral signature of the notch filter, showing how longer fibers improve resolving power.*

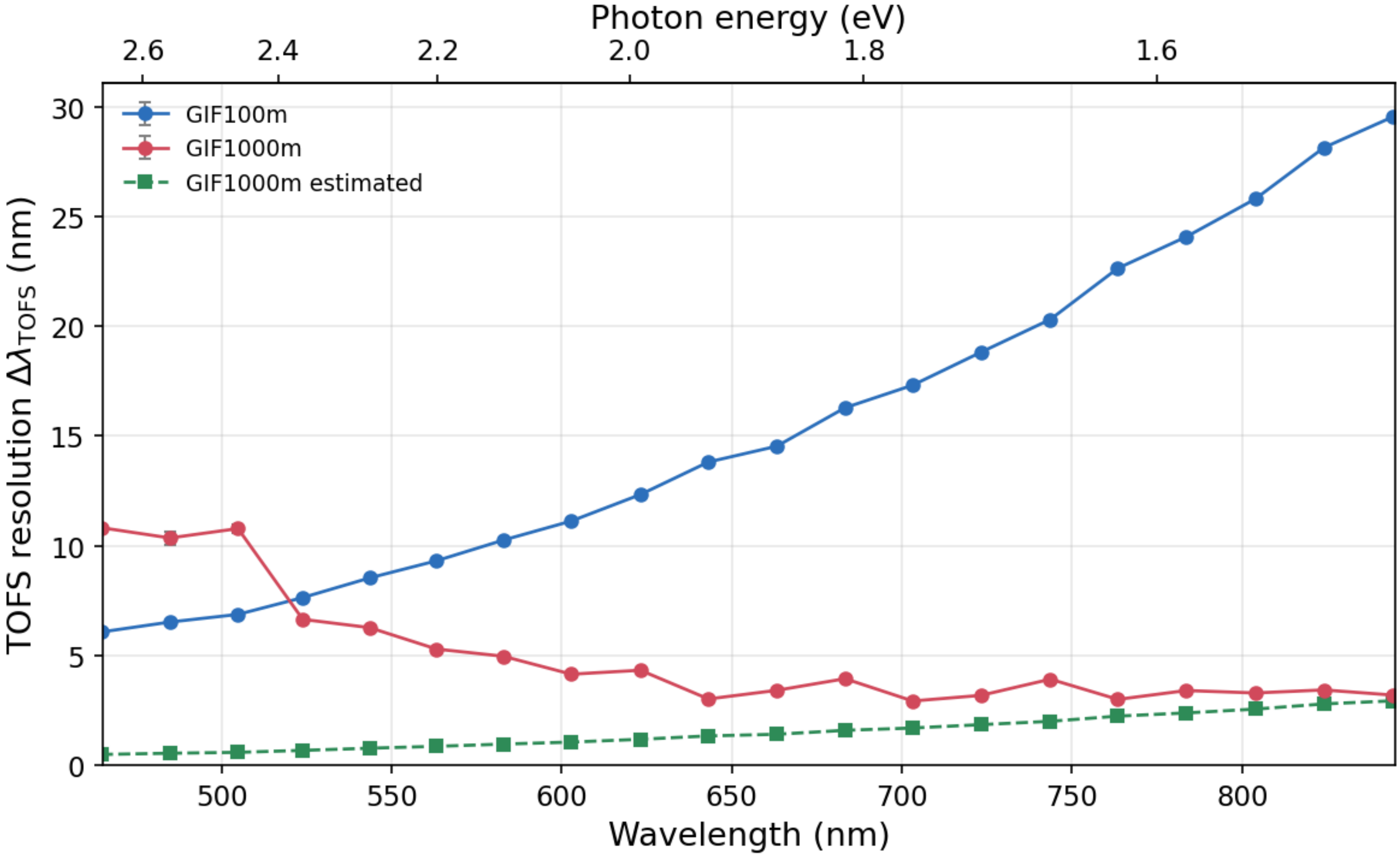

*Figure S3: Wavelength-dependent spectral resolution, $\Delta\lambda_{TOFS}(\lambda)$, of the TOFS for two graded-index fibers [Thorlabs GIF625-100 (blue) and GIF625-1000 (red)] measured with the wavelength tunable laser. The resolution is obtained from the measured temporal response of a light pulse of known color through the long fiber and by inverting the dispersion relation that maps photon arrival times to wavelength, $\Delta\lambda_{TOFS}(\lambda) = c\,FWHM_L(\lambda)/(L\mid dn_g/d\lambda\mid)$, where $FWHM_L(\lambda)$ is the full width at half maximum of the TOFS peak after propagating through the long fiber; $L$ is the fiber length; and $n_g(\lambda)$ is the group refractive index. Above $600\ nm$, we obtain an optical resolution of $(3.6 \pm 0.4)\ nm$ for the long fiber, here limited by the pulse-bandwidth of the excitation laser. The longer fiber has good spectral resolution at longer wavelengths, suggesting good utility of TOFS in the NIR. The shorter fiber has worse spectral resolution explaining the missing notch filter signature in* Figure S2*. We can estimate the true TOFS resolution of the $1000\ m$ fiber (green) via the measured $100\ m$ spectral resolution by taking $3.6\ nm$ as the broadening due to the source via $\Delta\lambda_{TOFS}^{1000} = \sqrt{(\lambda_{TOFS}^{100})^2 - 3.6^2}/10$. This leads to a spectral resolution spread for the $1000\ m$ TOFS, ranging from $0.5\ nm$ for a wavelength of $470\ nm$ to $2.9\ nm$ at $840\ nm$.*